# SpectraPlot.com: Integrated Spectroscopic Modeling of Atomic and Molecular Gases

Christopher S. Goldenstein[1,2], Victor A. Miller[1], R. Mitchell Spearrin[1,3], Christopher L. Strand[1,4]

[1]*SpectraPlot Ltd., Calabasas, CA*

[2]*Purdue University, School of Mechanical Engineering, West Lafayette, IN*

[3]*University of California Los Angeles, Los Angeles, CA*

[4]*Stanford University, Stanford, CA*

**Abstract**

SpectraPlot is a web-based application for simulating spectra of atomic and molecular gases. At the time this manuscript was written, SpectraPlot consisted of four primary tools for calculating: 1) atomic and molecular absorption spectra, 2) atomic and molecular emission spectra, 3) transition linestrengths, and 4) blackbody emission spectra. These tools currently employ the NIST ASD, HITRAN2012, and HITEMP2010 databases to perform line-by-line simulations of spectra. SpectraPlot employs a modular, integrated architecture, enabling multiple simulations across multiple databases and/or thermodynamic conditions to be visualized in a single interactive plot window. The primary objective of this paper is to describe the architecture and spectroscopic models employed by SpectraPlot in order to provide its users with the knowledge required to understand the capabilities and limitations of simulations performed using SpectraPlot. Further, this manuscript discusses the accuracy of several underlying approximations used to decrease computational time, namely, the use of far-wing cutoff criteria.

## 1. Introduction

*1.1. Background and Motivation*

The ability to calculate atomic and molecular absorption and emission spectra is important to countless fields in science and engineering (e.g., atmospheric and planetary science [1, 2], combustion and propulsion [3]). Such calculations require 1) a model that includes the equations required to describe the most pertinent physics and 2) a database containing all the spectroscopic parameters needed by the model. A relatively large number of spectroscopic databases have been developed to enable the spectra of atomic and molecular gases to be modeled (e.g., NIST ASD [4], HITRAN [5, 1], BT2 [6], HITEMP [7], GEISA [8, 9], CDSD [2]). While incredibly useful, such databases can be difficult for non-experts to use. To mitigate this, several researchers have developed software for calculating spectra as a function of thermodynamic conditions (i.e., temperature, pressure, composition) via various spectroscopic databases (e.g., spectralcalc.com [10], HITRAN on the Web [11], HITRAN Application Programming Interface (HAPI) [12]). SpectraPlot is a unique software tool due to it: 1) combining simulations of atomic and molecular spectra, 2) utilizing multiple databases, 3) employing an interactive web-based user interface, and 4) being open access. The

---

☆Corresponding author: csgoldenstein@purdue.edu

primary goal of SpectraPlot is to provide scientists, engineers, and students of all backgrounds with an easy-to-use and -access web-based application for simulating spectra of atomic and molecular gases. In this paper, we present the architecture, spectroscopic models, and assumptions employed by SpectraPlot in order to provide users with the knowledge required to understand the capabilities and limitations of simulations performed using SpectraPlot.

*1.2. Architecture*

SpectraPlot was designed to provide centralized access to spectroscopic data and integrated visualization of useful quantities derived from this data (e.g., absorption spectra). SpectraPlot has a modular architecture to enable processing of different data types and formats (e.g., HITRAN, NIST, etc.), as depicted in Figure 1. There is a single user interface for a given output (distinguished by tabbed headers at the top of each solver), and within that user interface, there are unique input fields, and an individual data processor and database for each data type. When a request is made to plot a selected data type, the data processor queries the appropriate database, computes the desired quantity, and outputs the result to a single output plot in the user interface. The single output enables comparison of different databases, a useful ability that can be cumbersome. This architecture works well for existing data types (i.e., catalogued, well-organized databases), and its simplicity allows for easy maintenance and updating. SpectraPlot is database-agnostic by design, and can be scaled to include a wide range of data formats.

SpectraPlot is written in Python (2.7) as a Flask application, and it is hosted on Amazon Web Services as an Elastic Beanstalk application. Databases are hosted on an Amazon RDS MySQL instance, and all spectroscopic simulations are run serially in Python on an Amazon EC2 instance. The user interface is built primarily using D3.js, an open source Javascript plotting library.

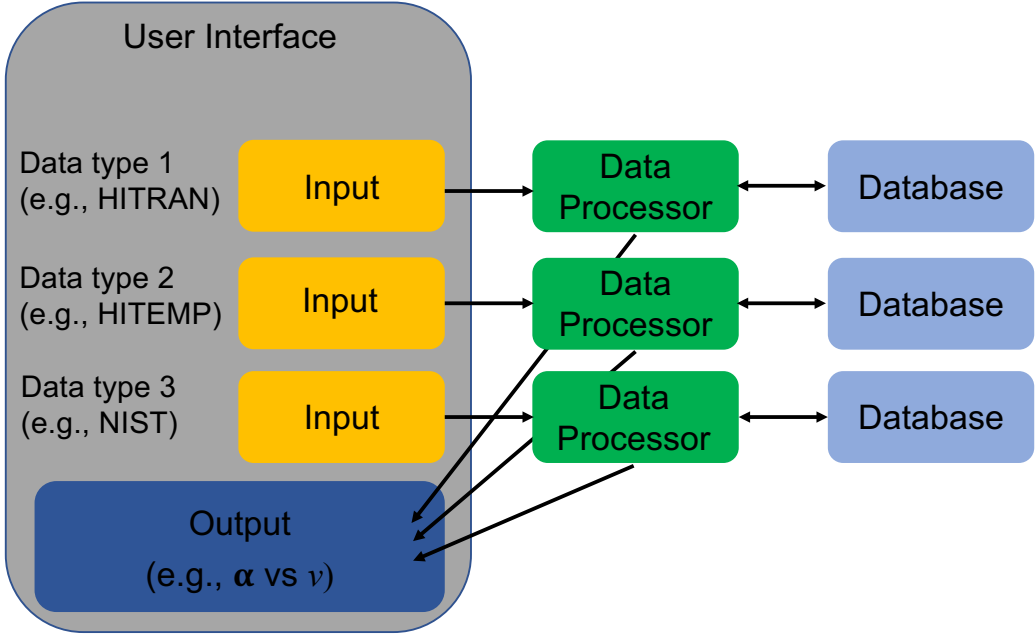

Figure 1: Data handling and visualization architecture employed by SpectraPlot.

A number of controls and warnings have been built into the user interface to ensure that SpectraPlot is stable and robust for a wide range of users, computers, internet browsers, and simulation requests. The following controls and warning are embedded within the interface. 1) If the simulation range and $\nu_{step}$ combine to yield an absorption or emission spectrum longer than 30,000 data points, $\nu_{step}$ is automatically increased such that the simulated spectrum consists of 30,000 data points. 2) Only the first (in order of increasing $\nu_o$) 10,000 lines are displayed in the line survey tool. Warning messages are automatically displayed whenever these limits are exceeded. The bottleneck in most SpectraPlot simulations is in rendering the simulation results to the interactive user interface (i.e., not in computing the simulations), although this is not always the case. Performing absorption and emission simulations using a relational database is relatively fast compared to reading the databases from local .csv files (e.g., roughly 5× faster), and none of the computation is performed on the local user's machine. However, because simulations are computed discretely (i.e., the wavelength range of the simulation is uniformly discretized and solved for), a relatively wide wavelength range results in a large number of data points that need to be rendered. Future versions of SpectraPlot may incorporate improved simulation and plotting routines by, for example, increasing and decreasing the resolution of simulations in regions of high- and low-curvature, respectively.

*1.3. Tools and User Interface*

SpectraPlot consists of four primary spectroscopic modeling tools for calculating: 1) atomic and molecular absorption spectra, 2) atomic and molecular emission spectra, 3) transition linestrengths, and 4) blackbody emission spectra. Given the simplicity of the latter, this paper will focus on the first three tools.

*1.3.1. Contents*

Currently, the absorption, emission, and line survey tools employ the majority, but not entirety, of the NIST ASD [4], HITEMP2010 [7], and HITRAN2012 [1] databases. These databases were chosen since they represent three of the most widely used and comprehensive databases enabling line-by-line simulations of spectra at user-specified thermodynamic conditions. Using these databases, SpectraPlot can simulate spectra of 78 atomic species in their first 3 ionization levels, and 37 molecular species with select isotopologues consistent with a given database (listed on HITRAN*online* [13]). These species are listed below in Table 1 and 2. It should be noted that the databases included in SpectraPlot are not entirely comprehensive and continue to evolve. The original manuscripts describing the databases (see [4, 7, 1]) employed by SpectraPlot should be consulted to determine the contents of each database (e.g., spectral coverage, included bands, energy levels, isotopologues etc.).

*1.3.2. Absorption and Emission Tools*

After selecting the database, the absorption and emission tools enable the user to simulate spectra of three species at a time, however each spectrum is calculated *independently*. When using the HITRAN and HITEMP databases, spectra can be calculated for a pure substance ($\chi_{rad} = 1$) or for a binary mixture of the radiating species in a bath gas of air ($\chi_{rad} < 1$). When using the NIST ASD database, the bath gas is effectively user-specified via the collisional-broadening parameters that are entered. The user must also select the wavelength or frequency range of interest, the frequency step size of the simulation, the thickness of (i.e., path length through) the gas sample, the mole fraction of the radiating (i.e., absorbing/emitting) species and the gas temperature and pressure. When using the HITRAN or HITEMP databases, all available isotopologues of the radiating species are included in the simulation. After the simulation is complete, the spectra are returned to the user's web browser and displayed in an interactive plot window equipped with a zoom subpanel (see Figure 2) and cursor for displaying local values. In the emission tool, the zoom subpanel can also be used to calculate the spectrally integrated emission intensity within the spectral window that is displayed. Up to three additional simulations (for a maximum of 9 total spectra) can be displayed in the plot window at a time. The user can save the data to a .csv file and/or save the plot as a .png file (an example plot output by SpectraPlot is shown in Figure 2).

Table 1: Atomic species contained in SpectraPlot.

| H  | He | Li | Be | B  | C  | N  | O  |
|----|----|----|----|----|----|----|----|
| F  | Ne | Na | Mg | Al | Si | P  | S  |
| Cl | Ar | K  | Ca | Sc | Ti | V  | Cr |
| Mn | Fe | Co | Ni | Cu | Zn | Ga | Ge |
| Br | Kr | Rb | Sr | Y  | Zr | Mo | Tc |
| Ru | Rh | Pd | Ag | Cd | In | Sn | Sb |
| Te | I  | Xe | Cs | Ba | Lu | Hf | Ta |
| W  | Ir | Au | Hg | Ti | Pb | Bi | Po |
| At | Rn | Nd | Pm | Sm | Eu | Gd | Tb |
| Dy | Ho | Er | Tm | Yb | Lu |    |    |

Table 2: Molecular species contained in SpectraPlot.

| $H_2O$ | $CO_2$ | $O_3$ | $N_2O$ | $CO$ | $CH_4$ | $O_2$ |
|---|---|---|---|---|---|---|
| $NO$ | $SO_2$ | $NO_2$ | $NH_3$ | $HNO_3$ | $OH$ | $HF$ |
| $HCl$ | $HBr$ | $HI$ | $ClO$ | $OCS$ | $H_2CO$ | $HOCl$ |
| $N_2$ | $HCN$ | $CH_3Cl$ | $H_2O_2$ | $C_2H_2$ | $C_2H_6$ | $PH_3$ |
| $COF_2$ | $H_2S$ | $HCOOH$ | $O$ | $NO^+$ | $HOBr$ | $C_2H_4$ |
| $CH_3OH$ | $CH_3CN$ | | | | | |

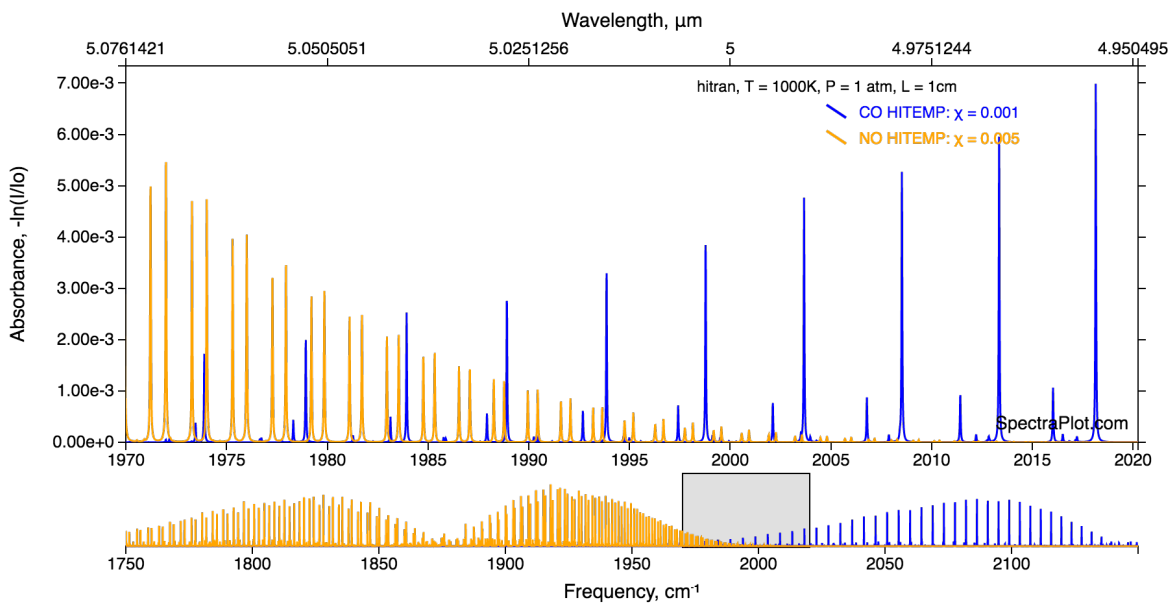

Figure 2: Example of simulated CO and NO absorbance spectra (calculated using the HITRAN2012 database [1]) and interactive plot window used by SpectraPlot.

*1.3.3. Linestrength Survey*

The linestrength survey tool enables users to survey the individual transitions within a given wavelength region. This is particularly advantageous when surveying large wavelength regions (i.e., on the order of microns) where calculations of spectra would be far more time consuming or when the spectroscopic parameters (linecenter, linestrength, etc.) are of explicit interest. The user must select the species and database of interest, the wavelength region, and the gas temperature, pressure, and mole fraction of the species of interest. The gas temperature is used to calculate the linestrength of each transition (see Section 2.2), the pressure is used to calculate the pressure-shifted linecenter of each transition (see Section 2.5), and the mole fraction is used to scale the linestrength values if desired. The latter of which is particularly useful when using the linestrength survey to predict the relative magnitude of spectra of multiple species that are in extremely different concentrations within a gaseous mixture. The user can also elect to impose a minimum linestrength to be displayed and whether or not to include isotopologues in the search results. The options for the latter are "all" or "none" where "none" refers to including only the most abundant isotopologue. The search results are then displayed in an interactive "stick-plot" (see Figure 3). By hovering the cursor over a given transition, the plot window will display values for the transition linecenter, linestrength, lower-state energy, air- and self-broadening coefficients and their temperature exponents (if available), and the isotopologue number when surveying the HITRAN and HITEMP databases.

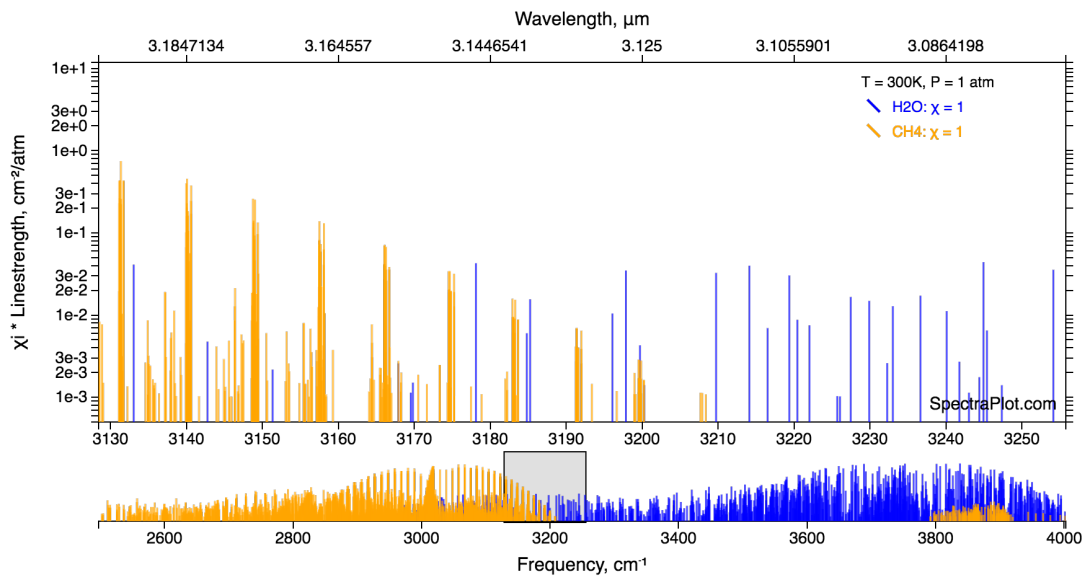

Figure 3: Linestrength of $H_2O$ and $CH_4$ transitions near 3 $\mu$m at 300 K. Linestrengths calculated as described in Sect: 2.2 using the HITRAN2012 database [1].

## 2. Spectroscopic Models

This section describes the most pertinent details required to understand the spectroscopic models employed by SpectraPlot.

## 2.1. Absorption and Emission Spectroscopy

SpectraPlot includes software for calculating absorption and emission spectra of gaseous species at a user-specified temperature, pressure, composition, and path length (i.e., gas slab thickness). All calculations are performed assuming a uniform gas sample and collimated light as depicted in Figure 4.

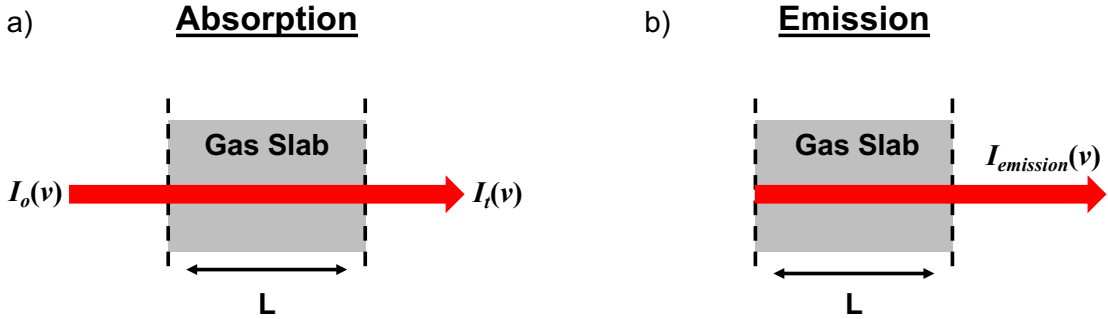

Figure 4: Schematic illustrating orientation between light and gas sample assumed in calculation of absorbance (a) and emission (b) spectra.

### 2.1.1. Calculation of Absorbance Spectra

The spectral absorbance, $\alpha$, is defined according to Beer's Law:

$$\alpha(\nu) = -\ln\left(\frac{I_t(\nu)}{I_o(\nu)}\right) \quad (1)$$

where $I_o$ and $I_t$ are the incident and transmitted light intensities at optical frequency $\nu$. For a uniform gas sample containing a single radiating (i.e., absorbing and emitting) species, the spectral absorbance can be related to gas properties using Eq. 2 and 3.

$$\alpha(\nu, T, P, \chi_{rad}, L) = k(\nu, T, P, \chi_{rad})L \quad (2)$$

$$k(\nu, T, P, \chi_{rad}) = \sum_j S_j(T) n \chi_{rad} \phi_j(\nu, T, P, \chi_{rad}) \quad (3)$$

Here $S_j(T)$ (cm$^{-1}$/molecule-cm$^{-2}$) is the linestrength of transition $j$ at temperature $T$ (K), $P$ (atm) is the gas pressure, $n$ (molecules/cm$^3$) is the number density of the gas (calculated using the ideal gas law), $\chi_{rad}$ is the mole fraction of the radiating species, $\phi_j$ (cm) is the transition lineshape function, and $L$ (cm) is the thickness of (i.e., path length through) the gas sample.

### 2.1.2. Calculation of Emission Spectra

The emissivity, $\varepsilon$, of a gas is defined according to Eq. 4:

$$\varepsilon(\nu, T, P, \chi_{rad}, L) = \frac{\widetilde{I}_{emission}(\nu, T, P, \chi_{rad}, L)}{\widetilde{I}_{BB}(\nu, T)} \quad (4)$$

where $\widetilde{I}_{emission}$ and $\widetilde{I}_{BB}$ are the spectral radiance of light emitted by the gas and a blackbody, respectively. The latter is calculated using Planck's Law. For a gas in thermal equilibrium, the emissivity equals the absorptivity of the gas (via Kirchoff's Law) and the spectral radiance emitted by a homogeneous gas slab of thickness $L$ can be calculated using Eq. 5.

$$\widetilde{I}_{emission}(\nu) = \widetilde{I}_{BB}(\nu,T)[1 - exp(-\alpha(\nu,T,P,\chi_{rad},L))] \quad (5)$$

Eq. 5 results from integrating the differential radiation balance across a gas slab of thickness $L$ [14] and implies that the emission spectrum of an optically thick ($\alpha \gg 1$) gas is equal to that of a blackbody. In SpectraPlot, $\widetilde{I}_{emission}$ and $\widetilde{I}_{BB}$ are calculated with units of power per solid angle per unit area per unit frequency or wavelength (e.g., W/sr-cm$^2$-cm$^{-1}$).

2.2. Calculation of Linestrengths

The linestrength of a given transition describes a molecule's propensity to absorb and emit light as it moves between two quantum states. This section discusses how the linestrength of atomic and molecular transitions are calculated using the parameters given in the HITEMP2010 [7], HITRAN2012 [1], and NIST ASD [4] databases. It is important to note that stimulated emission is accounted for in all equations used to calculated linestrengths in SpectraPlot.

2.2.1. HITRAN and HITEMP

The HITRAN and HITEMP databases list linestrengths at a reference temperature, $T_o$, of 296 K. In this case, the linestrength at $T$ is given by Eq. 6:

$$S_j(T) = S_j(T_o)\frac{Q(T_o)}{Q(T)}exp(\frac{-hcE_j''}{k}(\frac{1}{T} - \frac{1}{T_o}))(1 - exp(-\frac{hc\nu_{o,j}}{kT}))(1 - exp(-\frac{hc\nu_{o,j}}{kT_o}))^{-1} \quad (6)$$

where $Q$ is the partition function of the absorbing species, $E''$ (cm$^{-1}$) is the lower-state energy of the transition, $c$ is the speed of light (cm-s$^{-1}$), $k$ (J-K$^{-1}$) is the Boltzmann constant, and $h$ (J-s) is Planck's constant. Simulations performed using the HITRAN and HITEMP databases include all isotopologues (for the radiating species) included in the database. The reference linestrength for each isotopologue is scaled by its natural abundance as already done in the HITRAN2012 and HITEMP2010 databases [1].

In the linestrength survey tool, linestrengths calculated using the HITRAN and HITEMP databases are output with units of cm$^{-2}$-atm$^{-1}$ (i.e., on a pressure-normalized basis). Eq. 7 can be used to convert between units of cm$^{-2}$-atm$^{-1}$ and cm$^{-1}$/molecule-cm$^{-2}$ for an ideal gas [14].

$$S_j(T)[cm^{-2} - atm^{-1}] = \frac{S_j(T)[cm^{-1}/(molecule - cm^{-2})] \times 7.34 \times 10^{21}}{T[K]} \quad (7)$$

2.2.2. NIST ASD

The NIST ASD tabulates the Einstein-A coefficient ($A$), lower-state energy, and upper- and lower-state degeneracies ($g'$ and $g"$, respectively) of each electronic transition. With this data it is more convenient to use Eq. 8 to calculate the transition linestrength.

$$S_j(T_{elec}) = \frac{\lambda^2}{8\pi}F(T_{elec})A_j\frac{g'_j}{g"_j}(1 - exp(-hc\nu_{o,j}/kT_{elec}))/c \quad (8)$$

Here, $\lambda$ is the center wavelength of the transition (cm), $F$ is the fractional population in the absorbing state, and $T_{elec}$ (K) is the electronic temperature of the absorbing species. For simulations performed using the NIST ASD, $T_{elec}$ is an additional user input to enable more accurate simulations of spectra of atomic gases that are not in electronic-translational equilibrium. For a gas in thermal equilibrium, $T_{elec} = T$ where $T$ is the kinetic temperature used for calculating number density, Doppler broadening, and collisional broadening. $F$ is calculated according to Eq. 9:

$$F(T_{elec}) = g" exp(-hcE"/kT_{elec})/Q_{elec}(T_{elec}) \qquad (9)$$

where $Q_{elec}$ is the electronic partition function of the absorbing species. In some cases, the NIST ASD does not list values of $A$ for a given transition. In these cases, SpectraPlot assigns a default value of $A = 10^7$ s$^{-1}$ and, as such, extreme caution should be exercised when simulating the spectra corresponding to such transitions. Transitions with default values for $A$ assigned are included in simulations of absorbance and emission spectra, but not in the results returned by the line survey tool.

## 2.3. Calculation of Partition Function

### 2.3.1. HITRAN and HITEMP

The HITRAN *Global Data* [5, 1] contains total internal partition sums (calculated according to Fischer et al. [15]) for each species and isotopologue in the HITRAN and HITEMP database at temperatures from 70 to 3000 K. In SpectraPlot, the partition function of a given species is calculated at the user-specified temperature via linear interpolation or, if necessary, extrapolation using the HITRAN *Global Data*.

### 2.3.2. NIST ASD

The NIST ASD does not provide pre-calculated partition function sums. As a result, a lookup table containing the electronic partition function of all atomic species and ions listed in SpectraPlot was generated for temperatures up to 800 eV (1 K = 8.617x10$^{-5}$ eV). The electronic partition function was calculated using the degeneracies and energy levels listed in the NIST ASD via Eq. 10:

$$Q_{elect}(T_{elec}) = \sum_i g_i exp(-hcE_i/kT_{elec}) \qquad (10)$$

where $i$ denotes a given energy level. For a given SpectraPlot simulation, $Q_{elect}$ is calculated at the user-specified temperature via linear interpolation within the lookup table or, if necessary, extrapolation.

## 2.4. Calculation of Lineshapes

A wide variety of lineshape profiles have been developed to model transition lineshapes (e.g., Lorentzian, Gaussian, Voigt, Galatry, Rautian) and improved models that account for more complex collision physics continue to be developed [16, 17]. For computational simplicity and consistency with the majority of spectroscopic databases, SpectraPlot uses the Gaussian, Lorentzian, and Voigt Profiles (the latter via the algorithm developed by Mclean et al. [18] and Martin and Puerta [19]) to model transition lineshapes. The Gaussian and Lorentzian Profiles are used when the Lorentzian-to-Doppler ratio is $<$ 0.001 and $>$1000, respectively, and the Voigt Profile is used in all other cases. The Doppler full-width at half-maximum (FWHM), $\Delta\nu_{D,j}$, is calculated using Eq. 11:

$$\Delta\nu_{D,j} = 7.1623 \times 10^{-7} \nu_{o,j} \sqrt{T/M} \qquad (11)$$

where $M$ (g-mol$^{-1}$) is the molecular weight of the radiator. For simulations using the HITRAN or HITEMP databases, the collisional (i.e., Lorentzian) FWHM, $\Delta\nu_c$, is calculated according to Eq. 12:

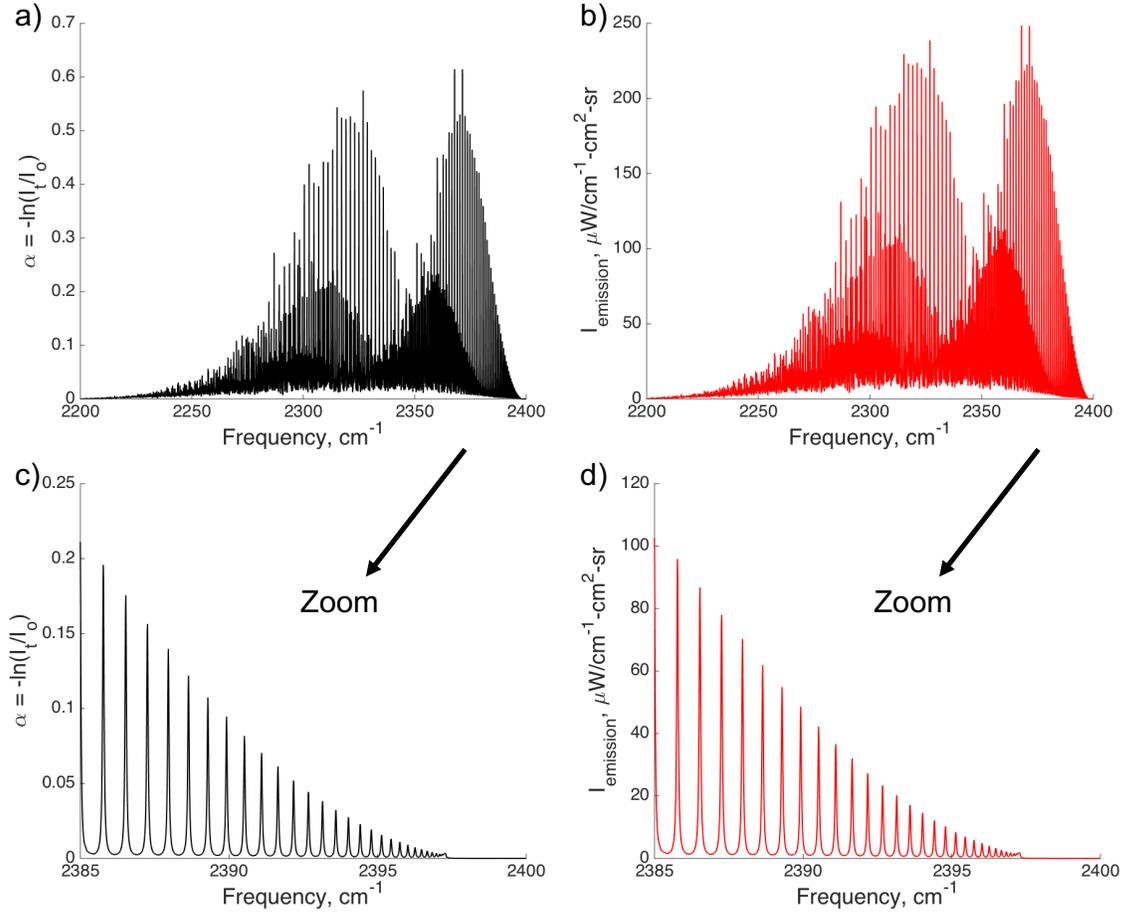

Figure 5: Simulated absorbance (a,c) and emission (b,d) spectra near 2300 cm$^{-1}$ for a 1 cm thick gas slab of 1% $CO_2$ in air at 1000 K and 1 atm. Simulation performed using SpectraPlot and the HITRAN2012 database [1].

$$\Delta\nu_{c,j} = 2P(\chi_{rad}\gamma_{self,j}(T) + (1-\chi_{rad})\gamma_{air,j}(T)) \quad (12)$$

Here, $\gamma_{self,j}$ and $\gamma_{air,j}$ are the collisional-broadening coefficients for radiator-radiator collisions (i.e., self broadening) and radiator-air collisions, respectively. At a given temperature, $\gamma$ is calculated using the power-law model (Eq. 13):

$$\gamma_j(T) = \gamma_j(T_o)(\frac{296}{T})^{n_j} \quad (13)$$

where $n_j$ is the perturber-specific collisional-broadening temperature exponent of transition $j$. For all species, an estimate ($n_{self}$ =0.75) for the self-broadening temperature exponent is used since HITRAN2012 and HITEMP2010 do not list temperature exponents for self-broadening coefficients. Figure 5 shows example absorbance and emission spectra of $CO_2$ near 4.3 $\mu$m calculated using the models described here.

The NIST ASD does not tabulate collisional-broadening coefficients. As a result, simulations of atomic spectra require the user to input a collisional-broadening coefficient and corresponding temperature exponent for the transition and gas mixture of interest. Figure 6 shows an example of simulated Na absorption spectra near 589 nm. The spectra were calculated using $\gamma(T_o) = 0.25$

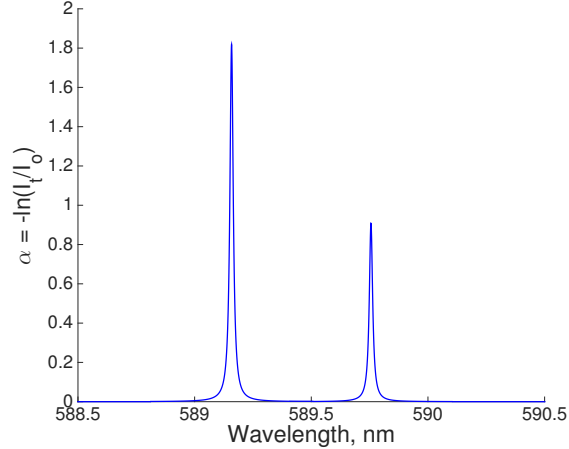

Figure 6: Simulated absorbance spectrum near 589 nm of 0.1 ppm of Na in $N_2$ at 296 K and 1 atm with a path length of 1 cm. Simulation performed using SpectraPlot and the NIST ASD database [4].

$cm^{-1}$-$atm^{-1}$ which is representative of these Na transitions dilute in $N_2$ [14].

*2.4.1. Far-Wing Cutoff*

Since transition lineshapes extend infinitely in frequency space, a rigorous calculation of absorption and emissions spectra would require simulating the lineshape of every transition in the database regardless of the wavelength range of interest. This is often computationally impractical (e.g., the HITEMP2010 database for $H_2O$ alone contains more than 1 billion transitions) and, further, rather futile given the inaccuracy of all Lorentzian-based lineshape models in the far wings [20]. To mitigate the former issue, SpectraPlot imposes a far-wing cutoff frequency to limit: 1) the number of lines included in a given simulation and 2) to reduce the wavelength range overwhich each transition is simulated. The cutoff frequency, $\Delta\nu_{cutoff}$, is determined using the user-specified pressure as follows: If $P < 10$ atm, then $\Delta\nu_{cutoff} = 200\ cm^{-1}$, else if $10$ atm $< P < 100$ atm then $\Delta\nu_{cutoff} = 20\ cm^{-1}/atm \times P$ and in all other cases, $\Delta\nu_{cutoff} = 2000\ cm^{-1}$. These simple criteria for determining $\Delta\nu_{cutoff}$ were determined by studying the spectra of $H_2O$ and $CO_2$ near 2.8 and 4.3 $\mu$m, respectively, at various temperatures and pressures. These species were chosen as ambassadors for the entire HITRAN2012 and HITEMP2010 databases due to the large number of users and, therefore, areas of science and engineering that rely on accurate simulations of these species' spectra, and 2) because these species exhibit markedly different band characteristics.

Using the far-wing cutoff criteria, each SpectraPlot simulation is performed as follows. All transitions with a linecenter frequency (in vacuum) located between $\nu_{start} - \Delta\nu_{cutoff}$ and $\nu_{end} + \Delta\nu_{cutoff}$ are included in the simulation where $\nu_{start}$ and $\nu_{end}$ are the user-specified bounds of the simulation. In addition, the lineshape of each transition included in the simulation is calculated from $\nu_o - \Delta\nu_{cutoff}$ to $\nu_o + \Delta\nu_{cutoff}$, and the lineshape of a given transition is set to zero at all frequencies outside this frequency range.

Figures 7 and 8 illustrate the error (i.e., the residual between spectra simulated with and without the far-wing cutoff) in $H_2O$ and $CO_2$ spectra, respectively, resulting from imposing the far-wing cutoff at pressures of 0.1 and 100 atm and temperatures of 296 and 2500 K. Figure 7 shows that implementing the far-wing cutoff introduces less than 0.1% error (quoted as percent of peak-band-absorbance) in the $H_2O$ spectra near 2.8 $\mu$m at the temperatures and pressures investigated. As expected, the error is generally larger for higher pressures (manifesting as a near-constant offset in the spectra, see Fig. 7c, 7d, 8c, and 8d) and smaller at higher temperatures due to the, typically, reduced collisional-broadening and narrower lines. In comparison, Figure 8 illustrates that the error in the $CO_2$ spectra near 4.4 $\mu$m is less than $10^{-4}$% of the peak-band absorbance at the conditions investigated. This smaller error primarily results from the fact that the $CO_2$ bands near 4.3 $\mu$m span a narrower frequency range compared to those of $H_2O$ near 2.8 $\mu$m. While larger *local errors* are certainly expected (particularly at frequencies located greater than $\Delta\nu_{cutoff}$ away from a strong band's edge), we argue that such errors will result from ignoring far-wing lineshape contributions that conventional Lorentzian/Voigt-based lineshape modeling cannot accurately account for regardless. Further, it is worth noting that these simulations were all performed using HITRAN2012 (as opposed to HITEMP2010) in order to achieve a manageable computational time. We expect that the residuals shown in Figures 7 and 8 would all increase modestly if the HITEMP2010 database were used in these calculations (due to the increase in opacity from high-energy transitions not included in HITRAN2012).

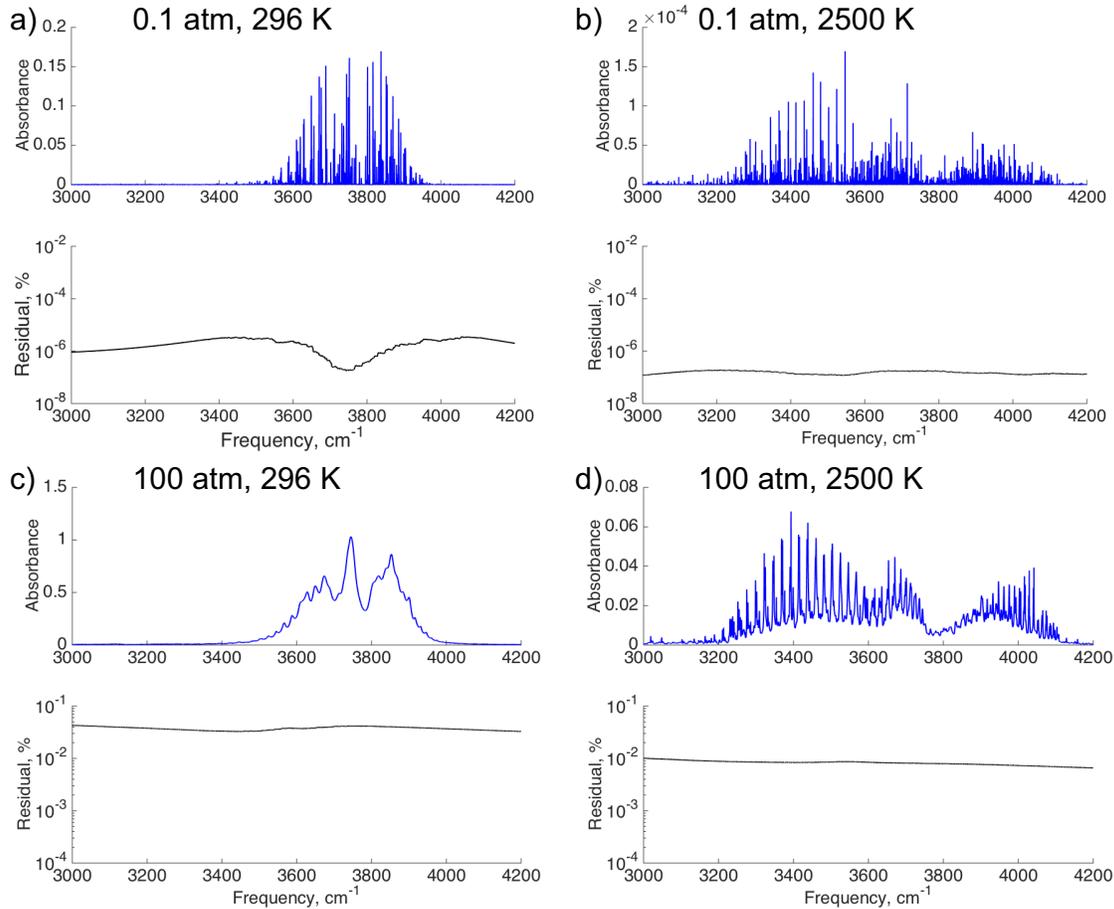

Figure 7: Peak-absorbance-normalized residual between $H_2O$ absorbance spectra near 3600 cm$^{-1}$ (2.78 $\mu$m) simulated using the HITRAN2012 database [1] with and without the far-wing cutoff imposed.

*2.5. Calculation of Transition Linecenter*

Transition linecenters are taken from the HITRAN2012 [1], HITEMP2010 [7] and NIST ASD [4] databases, which list each transition's linecenter frequency in vacuum ($\nu_{o,vac}$). Pressure shifting is accounted for in simulations performed using the HITRAN and HITEMP databases according to Eq. 14.

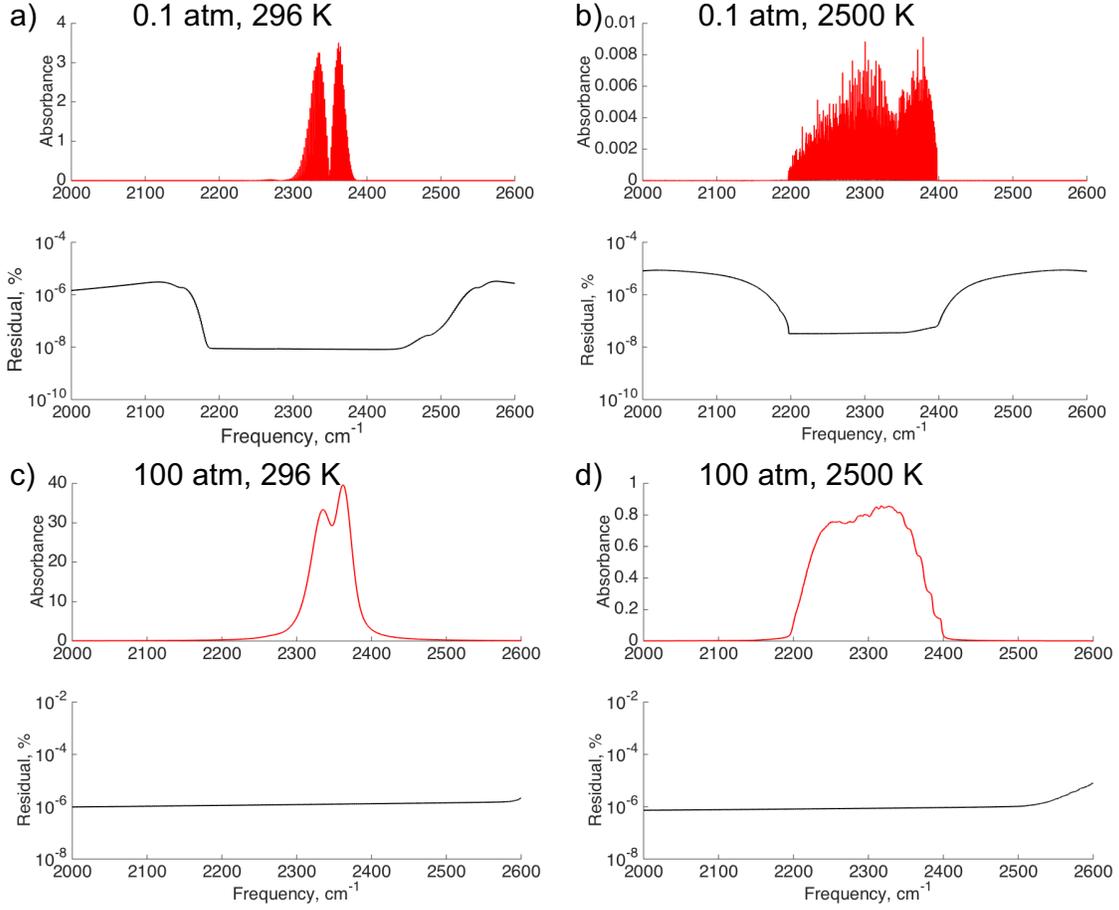

Figure 8: Peak-absorbance-normalized residual between $CO_2$ absorbance spectra near 2300 cm$^{-1}$ (4.35 $\mu$m) simulated using the HITRAN2012 database [1] with and without the far-wing cutoff imposed.

$$\nu_{o,j} = \nu_{o,vac,j} + P(1 - \chi_{rad})\delta_{air,j}(T) \quad (14)$$

Here, $\delta_{air}$ is the pressure-shift-coefficient for a given transition in air (tabulated in HITRAN and HITEMP at 296 K). The temperature dependence of $\delta_{air}$ is modeled using the power-law model (analogous to Eq. 13). However, since temperature exponents for $\delta_{air}$ are not listed in HITRAN2012 and HITEMP2010, an estimated value of 0.96 is used for all transitions following the recommendation put forth in [14]. Pressure-shifting is not accounted for in SpectraPlot simulations performed using the NIST ASD since pressure-shift coefficients are not given in the database.

## 3. Conclusions

SpectraPlot is a web-based application consisting of 4 primary tools for calculating: 1) atomic and molecular absorption spectra, 2) atomic and molecular emission spectra, 3) transition linestrengths, and 4) blackbody emission spectra. SpectraPlot uses the NIST ASD, HITEMP2010, and HITRAN2012 databases via a modular architecture that enables simulation results to be visualized across multiple databases, species, and thermodynamic conditions within a single, interactive user interface. Users can save simulation results as a .csv file and figures as a .png file.

This manuscript described the spectroscopic models and approximations used by SpectraPlot. Several approximations are employed to improve the robustness and computational efficiency of spectroscopic simulations, as well as to overcome deficiencies in the spectroscopic databases employed. Most significantly, SpectraPlot employs a far-wing cutoff frequency to limit the number of lines included in a given simulation and to reduce the wavelength range overwhich each transition is simulated. Simple criteria for determining the far-wing cutoff frequency were determined by simulating spectra of $H_2O$ and $CO_2$ over a broad range of temperatures and pressures. The simulation results suggest that the far-wing cutoff approximation introduces negligible error ($< 0.1\%$ of peak band absorbance) for a wide range of temperatures and pressures.

## 4. Acknowledgements

The authors would like to thank the many contributors to the NIST ASD, HITRAN, and HITEMP databases for making their data open access and enabling the inclusion of their content in SpectraPlot. Further, we thank our evolving list of corporate sponsors that help financially support SpectraPlot.com.